\documentclass[aps,prb,amsmath,amssymb,reprint,floatfix,superscriptaddress,longbibliography]{revtex4-2}
\usepackage{graphicx}
\usepackage{dcolumn}
\usepackage{wasysym}
\usepackage{bm}
\usepackage{color}
\usepackage{courier}

\begin{document}

\title{Synchronization-dissipation dynamics in the cardiorespiratory system}

\author{Joshua R. Border}
 \affiliation{Department of Physics, University of Bath, Bath BA2 7AY, United Kingdom}

\author{Alain Nogaret}
 \email{A.R.Nogaret@bath.ac.uk}
 \affiliation{Department of Physics, University of Bath, Bath BA2 7AY, United Kingdom}

\author{Andrew Lefevre}
 \affiliation{Department of Physics, University of Bath, Bath BA2 7AY, United Kingdom}

\author{Vishal Jain}
 \affiliation{Department of Physics, University of Bath, Bath BA2 7AY, United Kingdom}

\begin{abstract}
Dissipative coupling is known to induce synchronization.  Conversely it may be hypothesized that oscillators driven to synchronize may reduce power dissipation in their coupling.  The latter scenario is realized in the human cardiorespiratory system where cardiac and respiratory rhythms are controlled by the central nervous system while interacting viscoelastically through the pulmonary vasculature.  Here we examine the functional significance of this coupling which is observed in respiratory sinus arrhythmia (RSA).  By modelling electrical and viscoelastic interactions within the cardiorespiratory system, we identify the conditions leading to synchronization.  We demonstrate that, when present, synchronization reduces cardiac power losses by $10\%$ in humans and up to 55\% in other species.  The predicted gain in cardiac output is compared to the gain observed \textit{in-vivo} by pacing the heart with a device restoring RSA.  It is therefore surmised that RSA may improve cardiac pumping efficiency by reducing dynamic stress and power dissipation in the pulmonary vasculature.
\end{abstract}

\maketitle

\section{Introduction}

Synchronization-dissipation is the process by which self-sustaining oscillators synchronize via dissipative coupling~\cite{Hale1996,Lu2023}.  It occurs when the coupling dissipates energy at a rate that is greater than the difference in natural frequencies of the two oscillators.  Synchronization-dissipation was observed in many physical systems such as spin ensembles~\cite{Li2023}, optically coupled dimers and atoms~\cite{Zhu2024,Masson2022}, quantum bits~\cite{Cabot2020} and populations of biological oscillators~\cite{Winfree1967}.  In contrast, biological rhythms often synchronize through entrainment by an external signal~\cite{Glass2001}.  This signal is transmitted to the oscillator via a nonlinear pathway.  For example retinal cells entrain the circadian pacemaker via the retinal ganglion~\cite{Berson2002}, whereas lung stretch receptors modulate heart rate via brainstem central pattern generators~\cite{Smith2007}.  The focus of this paper is to examine the functional significance of cardiorespiratory synchronization and the possible reasons why it is pervasive across Evolution.  One manifestation of cardiorespiratory coupling is Respiratory Sinus Arrhythmia (RSA) whose functional significance has long been controversial~\cite{Anrep1936,BenTal2012,Giardino2003,Mortara1994,Skytioti2022,Hayano1996,Hirsch1981,Eckberg2003}.  In RSA, the respiratory rhythm modulates the cardiac rhythm by increasing heart rate during each breadth intake.  In addition, experiments on human athletes have shown that the mean heart frequency locks to the respiration frequency suggesting a degree of synchronization is also present in RSA~\cite{Schafer1999,Rzeczinski2002,Lotric2000,Kralemann2013,Iatsenko2013,Bartsch2007,Bartsch2012}.  Further progress towards understanding RSA has recently been made through a series of $in-vivo$ experiments that paced the heart with a nonlinear oscillator to artificially restore RSA coupling~\cite{OCallaghan2019,Shanks2022,Riesenhuber2023,OCallaghan2016}.  The device paced the sinoatrial node with a neuronal oscillator entrained by a lung inflation sensor to emulate the nonlinear adaptation of heart rate to respiration ~\cite{Nogaret2013,Pacemaker2021,Chauhan2021,Zhao2015}.  These experiments showed that restoring the RSA modulation of heart rate by respiration effectively increased cardiac output by 17\%-20\% over monotonic pacing at the same average heart rate~\cite{Shanks2022,OCallaghan2019}.

In this paper we show that RSA helps reduce the work of the right ventricle by reducing the viscoelastic losses of the oscillatory component of blood flow in the pulmonary vascular system.  Our calculations show the formation of regions of low energy dissipation that largely coincide with frequency domains where the phase of the cardiac oscillator is locked to the respiratory cycle.  In these domains a sequence of $n$ cardiac oscillations repeats every $m$ respiratory cycles and is stable against small changes in respiration frequency.  The domain bandwidth increases with the RSA dose.  The RSA dose is the difference in natural cardiac frequencies in the inspiratory and expiratory phases normalised by the frequency in the expiratory phase.  The domain bandwidth is also strongly modulated by the duration of the inspiratory phase.  In the $m=3:n=1$ domain~\cite{OCallaghan2019,Shanks2022}, which is relevant to human physiology, we found that cardiorespiratory coupling reduces the power dissipated in the pulmonary vasculature by $\approx10\%$ relative to monotonic pacing.  This power gain is consistent with the increase in cardiac output observed experimentally when RSA pacing is switched on.  This suggests that RSA might play an important role in increasing cardiac pumping efficiency by reducing the workload of the right ventricle.

\section{Mechanism for improved cardiac efficiency under RSA}

The right ventricle of the heart pumps blood into lungs through large arteries that branch out into a dense network of pulmonary capillaries.  Capillaries vessels are $7.0\pm1.5\mu$m in diameter~\cite{Short1996}.  They perfuse lung alveoli allowing oxygen uptake and release of carbon dioxide.  The capillary wall is made of a single layer of endothelial cells that can stretch by a considerable amount while minimizing its surface tension.  Lung inflation applies an axial strain to capillaries that is accompanied by a reduction in capillary diameter.  Each breadth intake thus increases the pulmonary vascular impedance.  The heart expends approximately 1/6th of its total power in the static impedance of the lung vasculature~\cite{Lammers2013}.  This is Poiseuille dissipation under laminar blood flow.  A further 40\% of right ventricle power is lost to the oscillatory component of blood flow~\cite{Milnor1966,Kovacs2012,Knudsen2018}.  This power is dissipated through viscoelastic losses in the capillary walls and the surrounding lung tissue during stretch-and-contraction cycles.  Oscillations in vessel size are driven both by cardiac and respiratory rhythms as follows.  During a breadth intake, capillary vessels experience a \textit{positive} axial strain ($\epsilon_R>0$) that can be as large as 35\%.  Conversely, when the heart contracts during systole, the rapid surge in blood pressure expands the capillary diameter producing a \textit{negative} axial strain ($\epsilon_H<0$).  This occurs as surface tension in the capillary wall works to keep the volume of the capillary constant.  Strain being additive, the total axial strain $\epsilon \equiv \epsilon_R(\Omega)+\epsilon_H(\omega)$ is clearly a function of the respiratory frequency $\Omega$, the cardiac frequency $\omega$, and their relative phase.

The viscoelastic power dissipated per unit volume of the capillary is easily obtained as $p_v=\eta\dot{\epsilon}^2$ where $\eta$ is the viscosity of the capillary wall (Supplementary Information 1).  The mean power dissipated over a cycle is therefore proportional to the variance of the rate of change of the axial strain $Var(\dot{\epsilon})$.  Expanding the variance of the difference $\dot{\epsilon}_R(\Omega)-|\dot{\epsilon}_H(\omega)|$, the mean viscoelastic power dissipation is:

\begin{equation}
\langle p_v \rangle = \eta \left\{ Var(\dot{\epsilon}_R)+Var(|\dot{\epsilon}_H|)-Cov(\dot{\epsilon}_R,|\dot{\epsilon}_H|) \right\}.
\label{eq:eq1}
\end{equation}

\noindent If the cardiac and respiratory rhythms are uncorrelated, the covariance in Eq.\ref{eq:eq1} is zero and the power dissipated is the sum of the variances of respiratory and cardiac strain rates taken independently.  If however correlation is present and the cardiac and respiratory strain rates, $|\dot{\epsilon}_H|$ and $\dot{\epsilon}_R$, oscillate in phase with each other, their covariance will be positive.  Mean power dissipation will thus be lower.  Phasic synchronization hence reduces dissipation by allowing the opposite strains $\epsilon_R$ and $\epsilon_H$ to compensate each other, producing a smaller total strain on the capillary in average.  With the same argument, anti-phasic synchronization would combine strains constructively leading to greater viscoelastic power dissipation.  Thus cardiorespiratory synchronization observed in RSA appears to be a plausible mechanism for minimizing the work done by the right ventricle.

To elaborate on this argument, we now build a model to compute the power gains brought by RSA.  This assumes that cardiac rhythm is both modulated by and synchronized to the respiratory rhythm on a beat-to-beat basis through the central nervous system.  One notes here that the oscillatory component of blood flow dissipates power through viscoelastic coupling only whereas the mean blood flow dissipates power through Poiseuille dissipation only.  This distinction has been discussed by Lammers et al.\cite{Lammers2013}.  One easily verifies this is true by comparing the power lost to Poiseuille and viscoelastic dissipation (Supplementary information 1).

\section{Method}

\begin{figure*}
\includegraphics[width=\linewidth]{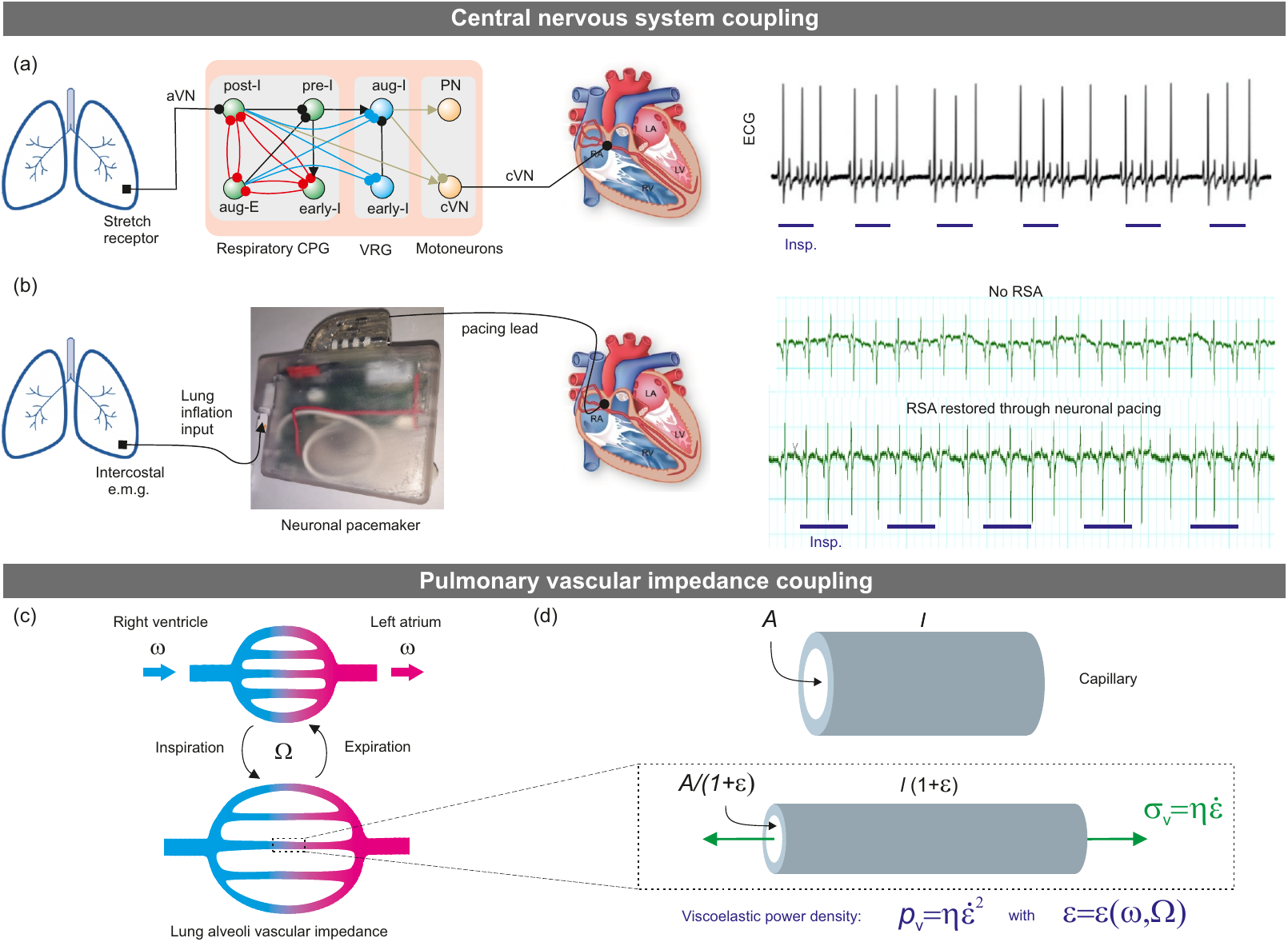} \\
\caption{\textbf{Cardiorespiratory coupling through the central nervous system and the pulmonary vascular impedance}
\\
(a) Central nervous system coupling.  Lung stretch receptors synchronize the respiratory rhythm, generated by the post-I, aug-E, early-E neuron populations, to the onset of inspiration.  The post-I neuron activation increases vagal tone (cVN) and slows down heart rate during the expiratory phase.  This produces the cardiac modulation (RSA) seen in a dog electrocardiogram.  (b) RSA restored by our neuronal pacemaker.  The pacemaker modulates heart rate within the respiratory cycle while synchronizing cardiac oscillations using the same nonlinear properties as central nervous circuits.  Electrocardiograms show the heart rate of an anaesthetized pig and with RSA artificially restored through neuronal pacing.  (c)  Viscoelastic coupling between cardiac and respiratory oscillators at the site of lung alveoli.  Pulmonary vascular impedance oscillates driven by the stretching and contraction of lung alveoli at frequency $\Omega$ and the systole/diastole cycling of blood pressure at cardiac frequency $\omega$.  (d) Viscoelastic power density $p_v$ dissipated by the capillary under axial strain $\epsilon$.  aVN$\equiv$ vagus nerve $a$ fibre, cVN$\equiv$ $c$ fibre, VRG$\equiv$ ventral respiratory group, e.m.g. $\equiv$ electromyography }
\label{fig:fig1}
\end{figure*}

Our model aims to compare the cardiac power dissipated under RSA modulation relative to monotonic pacing by accounting for electrical and mechanical interactions between heart and lungs.  These interactions are schematically depicted in Fig.\ref{fig:fig1}.  In vertebrates, RSA is generated in the central nervous system.  Central pattern generators in the brainstem pick up the onset of inspiration from lung stretch receptors.  They in turn increase vagal tone post-inspiration (cVN) to effectively slow down heart rate during the expiratory phase.  This action produces the heart modulation which is the signature of RSA (Fig.\ref{fig:fig1}a).  We model the brainstem-cVN-sinoatrial pacemaker complex with a neuronal oscillator.  A similar neuronal oscillator was implemented within the neuronal pacemaker one of us has developed~\cite{Nogaret2013,Pacemaker2021,Chauhan2021} and trialled \textit{in-vivo} to artificially restore RSA~\cite{OCallaghan2019,Riesenhuber2023,Shanks2022}.  The neuronal oscillator was stimulated by a rectangular current waveform that modulated the frequency of voltage oscillations.  The current injected during the expiratory phase $I_{exp}$ produced the base heart rate.  A larger current $I_{insp} > I_{exp}$ was injected during the inspiratory phase to produce an elevated heart rate.  The onset of inspiration was detected by an electromyography sensor placed on intercostal muscles or the diaphragm (Fig.\ref{fig:fig1}b).  The timed alternation of these currents produced the beat-to-beat modulation of cardiac frequency by respiration that characterises RSA.  The action potentials of the neuronal oscillator were used to stimulate the heart via a cardiac pacing lead.  This model also produced a nonlinear response to lung stretch receptors equivalent to that of brainstem circuits.  This is essential to replicate the synchronization of cardiac and respiratory rhythms observed experimentally~\cite{Schafer1999,Rzeczinski2002,Kralemann2013,Iatsenko2013}.  \textit{In-vivo} data show that the neuronal pacemaker, which implements the current model in hardware, has been effective in restoring cardiorespiratory coupling, increasing cardiac output in animal models of heart failure, and inducing cardiac remodelling after 3-4 weeks pacing~\cite{OCallaghan2019,Riesenhuber2023,Shanks2022}.

The action potentials setting the timing of electrical impulses, $t_i$, to the heart were synthesized by the Hodgkin-Huxley model~\cite{Hodgkin1952}:

\footnotesize
\begin{equation}
\begin{split}
C\frac{dV}{dt} & = g_{Na}m^3h(E_{Na}-V)+g_Kn^4(E_K-V)+g_L(E_L-V)+\frac{I(t)}{A} \\
\frac{dm}{dt} & = \frac{m_{\infty}(V)-m}{\tau_{m}(V)} \\
\frac{dh}{dt} & = \frac{h_{\infty}(V)-h}{\tau_{h}(V)} \\
\frac{dn}{dt} & = \frac{n_{\infty}(V)-n}{\tau_{n}(V)}
\end{split}
\label{eq:eq2}
\end{equation}
\normalsize

\noindent where $V$ is neuron output voltage; $m$, $h$ and $n$ are the gate variables of sodium and potassium ion channels. $I(t)$ is the current injected through soma area $A=2.9\times10^{-4}$cm$^2$.  $C=1.0\mu$F.cm$^{-2}$ is the membrane capacitance per unit area; $g_{Na}=69.0$mS.cm$^{-2}$, $g_K=6.9$mS.cm$^{-2}$ and $g_{Leak}=0.165$mS.cm$^{-2}$ are the areal conductance of the Na, K and Leak ion channels.  $E_{Na}=41.0$mV and $E_K=-100$mV are the reversal potentials.  In the steady state, gate activation / inactivation follow a sigmoidal dependence on $V$ as follows:

\begin{equation}
x_{\infty}(V)= 0.5 \left\{1 + \tanh \left[(V - V_{x})/dV_x\right] \right\}, \\
\label{eq:eq3}
\end{equation}

\noindent where $x\equiv \{m,h,n\}$ and the gate recovery times is:

\begin{equation}
\tau_{x}(V) = \tau_{0,x} + \epsilon_x \left\{ 1 - \tanh^2 \left[(V - V_{x}) / dVt_x \right] \right\},
\label{eq:eq4}
\end{equation}

\noindent Parameters were $V_m=-39.9$mV, $dV_m=10.0$mV, $dVt_m=23$mV, $t_{0m}=0.143$ms, $\epsilon_m=0.1$ms (Na$^+$ activation); $V_h=-65.4$mV, $dV_h=-17.0$mV, $dVt_h=27.2$mV, $t_{0h}=0.701$ms, $\epsilon_h=12.9$ms (Na$^+$ inactivation); $V_n=-34.6$mV, $dV_n=22.2$mV, $dVt_n=23.6$mV, $t_{0n}=1.29$ms, $\epsilon_n=4.31$ms (K$^+$ activation).  These parameters gave a resting voltage of -65.0mV and a current injection threshold of 0.52$\mu$A.

We stimulated the neuron above threshold with $I_{exp}=1.0\mu$A to produce a base heart rate $f_{exp}$.  During the inspiratory phase, the injected current increased to $I_{insp}=I_{exp}(1+rsa)$ to produce a higher cardiac frequency $f_{insp}>f_{exp}$. $rsa>0$ was the adjustable parameter used to set the RSA dose applied to the heart.  Introducing $\tau_k$ as the onset of the inspiratory phase in breathing cycles $k=0,\dots,N$ and $\beta$ ($0<\beta<1$) as the duty cycle of the inspiratory phase, the current waveform injected in the neuron was a series of square current steps given by:

\footnotesize
\begin{equation}
I(t)=\sum_{k=0}^{N-1} I_{exp}\left(1+rsa*\left[H(t-\beta\tau_{k+1}-(1-\beta)\tau_k)-H(t-\tau_k) \right]\right)
\label{eq:eq5}
\end{equation}
\normalsize

\noindent where $H()$ is the Heavyside step function.  In biology, the duty cycle of the inspiratory phase is set by the respiratory central pattern generator (Fig.\ref{fig:fig1}a).  This circuit has three neurons: Early-I, Post-I, Aug-E that discharge sequentially via mutually inhibitory synapses (red links) to produce the respiratory rhythm.  It is the firing delay between Early-I and Post-I neurons that determines the duration of the inspiratory phase hence $\beta$.  The value of $\beta$ is therefore subject specific and depends of the physiological state.  We integrated the current waveform $I(t)$ using a fifth order adaptive step size Runge-Kutta to generate the voltage oscillations of the neuron.  From this we extracted the timings of action potentials $t_i$, $i=0,\dots,n$ that display both RSA frequency modulation and synchronization to lung inflation.

We model viscoelastic coupling between cardiac and respiratory rhythms as schematically depicted in Fig.\ref{fig:fig1}c.  The viscoelastic power dissipated through stretching-contraction cycles is the combination of positive axial strain applied by the alveoli during lung inflation and a negative axial strain arising from the surge in blood pressure during systole.  We model the time dependence of the blood pressure-induced strain as a series of Gaussian peaks centered on the timings of heart beats $t_i$ calculated from Eqs.\ref{eq:eq2}-\ref{eq:eq4}.

\begin{equation}
\epsilon_H(t) = -\epsilon_H^0 \sum_{i=0}^{n} \exp\left(-\frac{(t-t_i)^2}{\Gamma^2} \right),
\label{eq:eq6}
\end{equation}

\noindent where $\epsilon_H^0>0$ is the cardiac oscillation amplitude and $\Gamma$ is half the duration of systole.  Systole lasts 0.3s in humans at rest hence we take $\Gamma=0.15$s.  We modelled the time dependence of the axial strain produced by lung inflation as a series of steps centered on the inspiratory phase with broadened edges to account for the progressive increase/decrease in respiratory flow:

\footnotesize
\begin{equation}
\epsilon_R(t) = \frac{\epsilon_H^0}{2} \sum_{k=0}^{N-1} \left\{ \tanh\left(\frac{t-\tau_k}{\delta}\right) - \tanh\left(\frac{t-(\beta\tau_{k+1}+(1-\beta)\tau_k)}{\delta}\right) \right\}
\label{eq:eq7}
\end{equation}
\normalsize

\noindent where $\delta$ is duration of tidal flow and $\epsilon_H^0$ is the respiratory oscillation amplitude.  For the sake of our proof of concept we assume that the breathing period is constant and equal to $T$.  Hence $\tau_k=kT$.  The duration of tidal flow measured by chest impedance recordings is typically $1/6^{th}$ of the respiratory cycle, giving $\delta=\pi/(3\Omega)$.  The total strain experienced by the capillary is thus $\epsilon(t)=\epsilon_R(t) + \epsilon_H(t)$.  Individual strain waveforms are plotted in Fig.S1.

To quantify the change in power dissipation under RSA pacing relative to monotonic pacing we compute the relative power loss as:

\begin{equation}
\frac{\Delta p}{p}=\frac{\int_0^{\tau_N} dt \left(\dot{\epsilon}_{RSA}^2-\dot{\epsilon}_{mono}^2 \right)}{\int_0^{\tau_N} dt \; \dot{\epsilon}_{mono}}.
\label{eq:eq8}
\end{equation}

\noindent Both $\epsilon_{RSA}$ and $\epsilon_{mono}$ are total strains calculated from Eqs.\ref{eq:eq6} and \ref{eq:eq7}.  However, in the former, the timings of cardiac pulses $t_i$ (Eq.\ref{eq:eq6}) are generated by the neuronal oscillator stimulated by the square wave current.  In the latter, the timings of cardiac contraction, $t_i^{mono}$, are equally spaced and given by $t_i^{mono}=(t_n-t_0)(i/n)+t_0$.  Here $(t_n-t_0)/n$ is the mean R-R interval averaged over $N$ respiratory cycles.  Our simulation used $N=128$ cycles.  The $n$ index labels the last cardiac pulse in this series.  The average cardiac frequency is then given by $\omega = 2\pi n/(t_n-t_0)$.  The relative power loss in Eq.\ref{eq:eq8} conveniently eliminates coupling constants (such as the capillary wall viscosity $\nu$) and complex geometrical aspects of the lung vasculature which remain the same under both RSA and monotonic pacing.  The only parameters affecting cardiorespiratory coupling are the amplitudes of the respiratory and cardiac strains, $\epsilon_R^0$, $\epsilon_H^0$, the duration of systole $\Gamma$ and of the inhalation/exhalation transient $\delta$.  Parameters $\Gamma$ and $\delta$ determine how the $\epsilon_H(t)$ and $\epsilon_R(t)$ waveforms overlap under the integral in Eq.\ref{eq:eq8}.  These parameters alter the magnitude of power dissipation but not cardiorespiratory synchronization that depends on the $rsa$ and $\beta$ parameters.  To first order, the absolute value of strain amplitudes $\epsilon_H^0$ and $\epsilon_R^0$ is relatively unimportant.  It is their ratio that enters the pre-factor of Eq.\ref{eq:eq8} hence affects the magnitude of power dissipation.  Our simulation used $\epsilon_R^0=0.25$ and $\epsilon_H^0=0.25$ which broadly correspond to the maximum strain on capillaries under normal breathing and systole respectively.  The relative power dissipation given by Eq.\ref{eq:eq8} may thus be compared to the cardiac output measured under RSA pacing by the device of Fig.\ref{fig:fig1}b relative to monotonic pacing by a conventional pacemaker~\cite{Shanks2022,OCallaghan2019}.

\section{Results}

\begin{figure*}
\includegraphics[width=\linewidth]{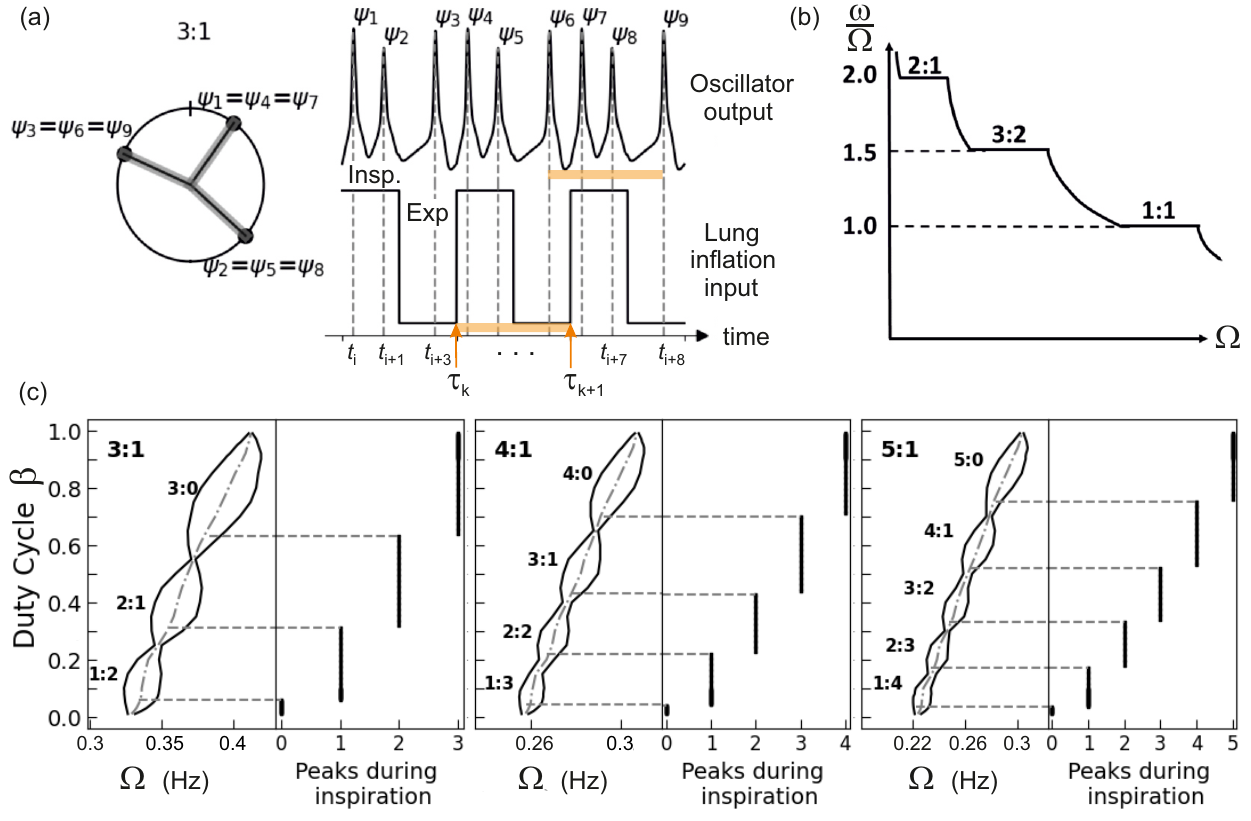} \\
\caption{\textbf{Cardiac entrainment by respiration: modulation with synchronization}
\\
(a) Cardiac rhythm entrained by the lung inflation signal.  The pacemaker drives the heart at a faster rate during a breadth intake than during the expiratory phase producing RSA.  The cardiac rhythm may also synchronize to respiration in which case the phase of cardiac oscillations $\psi_1$, $\psi_2$, $\psi_3$ locks to the respiratory cycle.  (b) Synchronization plateaux in the dependence of $\omega/\Omega$ on breathing frequency $\Omega$.  Each synchronization region $m:n$ has $m$ cardiac oscillations matching $n$ respiratory cycles.  (c) Synchronization domains $3:1$, $4:1$ and $5:1$ and their dependence on the inspiratory duty cycle $\beta$.  The troughs corresponds to one cardiac peak moving from the expiratory interval to the inspiratory interval.}
\label{fig:fig2}
\end{figure*}

The output of the neuronal pacemaker entrained by the lung inflation signal is shown in Fig.\ref{fig:fig2}a.  Pacemaker oscillations display 3 action potentials per breadth which is the typical 3:1 cardiac to breathing frequency ratio in humans at rest.  Cardiac oscillations are faster in the inspiratory phase than in the expiratory phase due to the higher injected current ($I_{insp}>I_{exp}$).  Hence neuronal pacing clearly restores RSA.  The greater $I_{insp}$ relative to $I_{exp}$ the greater the difference in cardiac frequencies $f_{insp}-f_{exp}$.  This example also shows phase synchronization occurring between the oscillator output and the respiratory drive: the phase of the 3 action potentials $\psi_1$,$\psi_2$ and $\psi_3$ repeats identically (modulo $2\pi$) from one respiration period to the next.  This means that the time interval between one action potential and the corresponding action potential in the previous breathing cycle is exactly the duration of the breathing period $T$.  This is shown by the orange stripe in Fig.\ref{fig:fig2} that spans intervals $t_{i+8}-t_{i+5}$ and $\tau_{k+1}-\tau_{k}$ that are equal in duration.  The nonlinearity of the neuronal oscillator biasses the mean cardiac frequency $\omega$ to maintain a constant $3:1$ ratio to the respiratory frequency as $\Omega$ changes.  This frequency locking only occurs within finite frequency bands (Arnold tongues).  The stronger the coupling between the entraining signal and the neuronal oscillator, the wider the Arnold tongues.  In our system, the coupling strength is the RSA dose, set by the $rsa$ parameter in Eq.\ref{eq:eq5}.

We computed the ratio $\omega/\Omega$ as a function of $\Omega$ in Fig.\ref{fig:fig2}b for $rsa=50\%$ and inspiratory duty cycle $\beta=0.4$.  The plot shows plateaux where a sequence of $m$ cardiac pulses synchronize to $n$ breathing intervals.  Within each $m:n$ plateau, the mean cardiac frequency latches on the breathing frequency until the detuning between the respiratory frequency and the natural frequency of the cardiac oscillator becomes greater than the strength of RSA coupling.  At this point, the phase of action potentials $\psi$ changes very rapidly upon small changes in $\Omega$.  The ratio $\omega/\Omega$ then decreases towards the next synchronization plateau.  Our simulations show that a sequence of neuronal oscillations can mode-lock to $n=1-3$ consecutive breathing periods.  Synchronization to $n>3$ breathing cycles was too faint to detect.

The width of $m:n$ plateaux is modulated by the duty cycle of respiration $\beta$.  This is shown in Fig.\ref{fig:fig2}c where one first considers the $3:1$ plateau and tracks the edges of this plateau as $\beta$ increases from 0 to 1.  Initially, the 3 action potentials present within the respiratory cycle will all belong to the expiratory interval, $(0_i,3_e)$.  When $\beta$ increases sufficiently, one of the 3 action potentials will move into the inspiratory interval leaving 2 in the expiratory interval, giving a $(1_i,2_e)$ pulse distribution.  Subsequent increases in $\beta$ will add a second action potential to the inspiratory interval $(2_i,1_e)$ and then a third $(3_i,0_e)$.  Every time an action potential transits from the expiratory region into the inspiratory region, the $3:1$ Arnold tongue exhibits a trough.  This transitory state clearly weakens cardiorespiratory synchronization.  Pulse patterns that are the most stable include action potentials located away from the transitions between inspiratory and expiratory states.  These stable states correspond to the bellies of Arnold tongues.  A similar pattern of bellies and troughs is seen in the $4:1$ and $5:1$ Arnold tongues as $\beta$ varies.  This allows us to conclude that the number of stable regions (bellies) is equal to the number of action potentials ($n$) in the repeating pulse sequence, while the number of unstable intermediary states (troughs) is $n+1$.  Fig.\ref{fig:fig2}c thus demonstrates that the inspiratory duty cycle plays a significant role in modulating the strength of cardiorespiratory synchronization.

\begin{figure*}
\includegraphics[width=\linewidth]{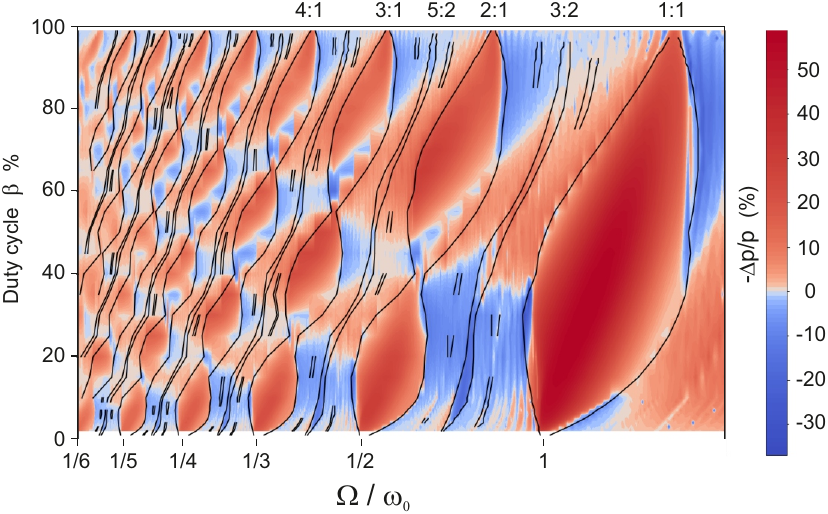} \\
\caption{\textbf{Viscoelastic energy gains mapped onto regions of cardiorespiratory synchronization}
\\
The colour map plots the difference in viscoelastic power dissipated by the heart paced with RSA relative to no RSA normalised by the latter.  The boundaries of synchronization regions (black lines) are superimposed to show the dependence of $m:n$ synchronization on the breathing frequency $\Omega$ and on the inspiratory duty cycle $\beta$.  $\omega_0$ is the cardiac frequency with no RSA modulation applied ($\beta=0$).  Parameters: $rsa=100\%$, $\epsilon_H^0=0.25$, $\epsilon_R^0=0.25$, $\Gamma=0.15$s, $\delta=\pi/(3\Omega)$.
}
\label{fig:fig3}
\end{figure*}

Fig.\ref{fig:fig3} superimposes the $m:n$ synchronization regions on a colour map of power gains generated by RSA pacing.  The power map was computed with Eq.\ref{eq:eq8} whereas the boundaries of synchronization regions were extracted from the extrema in the second derivative of the $\omega/\Omega$ traces with respect to $\Omega$ (Fig.\ref{fig:fig2}b).  On the $x$-axis, the respiration frequency increased in units of the base heart frequency, $\omega_0 = \omega(\beta=0)$.  As the inspiratory duty cycle increases (left axis of Fig.\ref{fig:fig3}), the average current stimulation per cycle increases hence the mean cardiac frequency also increases making $\omega>\omega_0$.  This explain why Arnold tongues shift to higher values of $\Omega/\omega_0$ as $\beta$ increases.  The respiration frequency must increase to maintain $m:n$ synchronization as the duty cycle increases.  Besides mode locking occurring within one breathing period, Fig.\ref{fig:fig3} clearly show the prevalence of mode locking to two consecutive periods: Arnold tongues $3:2,\dots,17:2$ are all observed within range albeit over a narrower frequency band than $m:1$ bands (Supplementary information 3).  Higher order model locking is less easily traceable but appears to be stable near the troughs of the $m:1$ Arnold tongues.

The power map shows that regions of cardiorespiratory synchronization largely coincide with reduced power dissipation.  The stronger the synchronization (the wider the Arnold tongue), the  greater the power saving that RSA pacing introduces.  For the $rsa=100\%$ coupling we use here, the maximum power savings at the centre of the $1:1$ region is $55\%$.  It is $10\%$ at the centre of the $3:1$ region corresponding to the human cardiorespiratory ratio at rest.  In the blue areas outside synchronization regions, RSA pacing produces dissipation similar or a few percent higher than under monotonic pacing.  In these regions, the rapid change in the phase of neuronal oscillations de-correlates them from the onset of breathing.  As a result the covariance $Cov(\epsilon_R,|\epsilon_H|)$ of Eq.\ref{eq:eq1} vanishes and the power dissipated becomes comparable to or slighter greater than under monotonic pacing that is also de-correlated from respiration.  Thus these data show that cardiorespiratory synchronization under RSA pacing can effectively increase the pumping efficiency of the heart by minimising the viscoelastic energy expended in the pulmonary vascular impedance.  This energy efficiency gain progressively decreases as the $n:m$ ratio increases and the synchronized state occupies narrower frequency bands.  Energy savings decrease from $55\%$ at $1:1$ to $\approx 4\%$ at $8:1$ (Fig.\ref{fig:fig3}).

\begin{figure*}
\includegraphics[width=\linewidth]{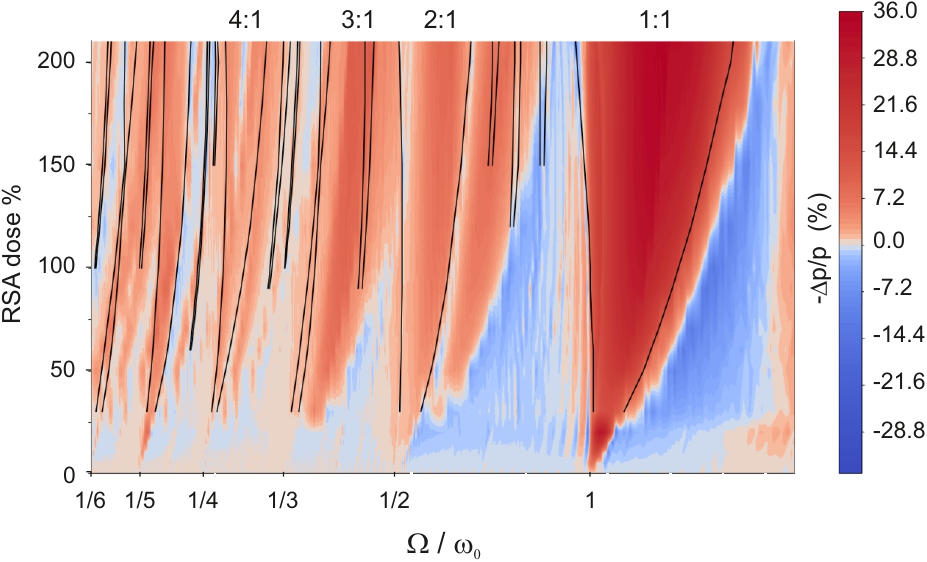} \\
\caption{\textbf{Spectral distribution of power gains and its dependence on RSA dose (\textit{rsa})}
\\
The bandwidth of $m:n$ synchronization regions increases with the RSA dose.  The energy gains within these regions (red domains) increase and saturate when RSA dose approaches 50\%.  In the blue domain, outside $m:n$ synchronization regions, the power dissipated under RSA pacing is marginally greater than under monotonic pacing.  $\beta=0.4$.}
\label{fig:fig4}
\end{figure*}

Increasing the coupling strength via the $rsa$ parameter has the effect of broadening Arnold tongues as seen in Fig.\ref{fig:fig4}.  The stronger coupling also stabilizes weaker patterns of synchronization that repeat every 2 or 3 consecutive respiratory cycles.  This is seen through the emergence of the $3:2$ and $5:2$ frequency domains as well as $4:3$ and $5:3$ omains at RSA doses greater than $100\%$ (see also Supplementary information 3).  Power gains are broadly centered on synchronization regions or clusters of such regions involving a principal mode such as $2:1$ with additional contributions from secondary modes such as $3:2$, $4:3$ and $5:3$.  Power gains tend to increase with RSA dose within each synchronized mode.  This increase is most rapid in the $1:1$ mode.  Then power gains saturate at $\approx 50\%$ RSA.  Power gains emerge more slowly in the higher order modes.

\begin{figure}
\includegraphics[width=\linewidth]{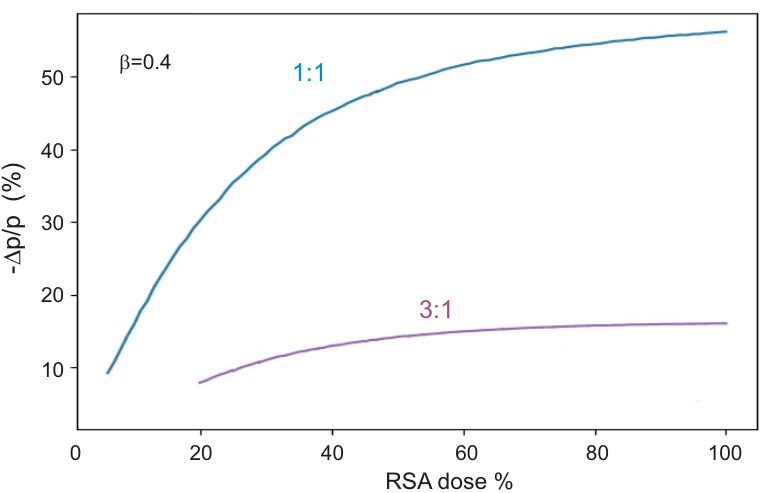} \\
\caption{\textbf{Dependence of power gains at the centre of a synchronization domain on RSA dose}
\\
Dependence of power gains produced by RSA inside the $1:1$ (blue trace) and $3:1$ Arnold tongues (purple trace).}
\label{fig:fig5}
\end{figure}

The increase in power gain at the centre of the $1:1$ and $3:1$ bands is shown in Fig.\ref{fig:fig5}.  Shanks at al. ~\cite{Shanks2022} applied $\approx 20\%$ RSA in pacing sheep with the device of Fig.\ref{fig:fig1}b implementing the present pacing model.  They observed an approximately $20\%$ increase in cardiac output.   Fig.\ref{fig:fig5} estimates a $10\%$ reduction in viscoelastic dissipation of the power generated by the right ventricle at a $20\%$ RSA dose.  Hence an increase in cardiac pumping efficiency of the same order might be expected for the whole heart in the 3:1 band.  This simulation therefore establishes that RSA facilitates the work of the right ventricle in pumping blood through the pulmonary vascular system by minimizing viscoelastic dissipation.  It is thus highly likely that improved right ventricle function resulting from this mechanism would improve the cardiac output as observed by Shanks et al.~\cite{Shanks2022}.

\section{Discussion}

The present simulation aimed to explain the gains in cardiac efficiency observed in animal models of heart failure paced under artificially restored RSA~\cite{OCallaghan2019,Riesenhuber2023,Shanks2022}.  The neuronal pacemaker implemented the nonlinear oscillator of Eqs.\ref{eq:eq1}-\ref{eq:eq4}~\cite{Pacemaker2021,Chauhan2021,Zhao2015,OCallaghan2016,Nogaret2013}.  This oscillator incorporates two key properties of the brainstem circuits responsible for modulating heart rate~\cite{Smith2007}.  Firstly, the neuronal oscillator restores heart rate variability within the respiratory cycle, resulting in shorter R-R intervals during the inspiratory phase than the expiratory phase which is the signature of RSA~\cite{Anrep1936}.  Secondly, the neuronal oscillator produces a nonlinear response to the lung inflation signal, similarly to neuronal circuits in the brainstem.  Nonlinear coupling biasses the mean heart rate away from its natural frequency by synchronizing the phase of cardiac oscillations to breadth intakes.  This results in the $m:n$ synchronization observed in human subjects~\cite{Schafer1999,Rzeczinski2002,Lotric2000,Kralemann2013}.  Here we further show how the strength of $m:n$ synchronization increases when the natural heart rate in the inspiratory phase increases relative to the heart rate in the expiratory phase.  The $rsa$ parameter increases the coupling of the cardiac oscillator to the driving force.  This has the effect of widening the $m:n$ frequency bands and stabilizing pulse patterns that repeat every 2 and 3 breathing cycles.  Varying the duty cycle of the inspiratory phase was found to modulate the width of $m:n$ frequency bands.  Synchronization is weaker at values of $\beta$ where the distribution of cardiac pulses among the inspiratory and expiratory intervals is unstable as it is most sensitive to a perturbation in $\beta$.

The power maps we have obtained in Figs.\ref{fig:fig3}-\ref{fig:fig4} show that cardiac synchronization under RSA pacing minimizes viscoelastic dissipation in the pulmonary vascular network.  By correlating cardiac systole with breadth intakes, phasic synchronization reduces the variance of their combined strains hence power losses (Eq.\ref{eq:eq1}).  Our prediction of a $\approx10\%$ lower dissipation in the $3:1$ mode at $rsa=20\%$ would translate in a comparable increase in output power $\dot{P}V$ by the right ventricle.  This is comparable to the $20\%$ overall increase in cardiac output measured under the same pacing settings.  It is therefore likely that one of the functions of RSA is to improve the efficiency of the right ventricle by minimizing dissipation in the pulmonary vasculature.  It is well-known that the pulmonary circuit has developed significant adaptations over Evolution to optimize hydraulic power efficiency, notably by operating at lower blood pressures than in systemic circuit~\cite{Milnor1966}, and because small improvements in right ventricle pumping efficiency disproportionately improve the overall power efficiency of the heart~\cite{Lammers2013}.

Our first-principle calculations thus show that RSA minimises the work done by the heart.  The implication is that loss of RSA modulation, an early prognosis of heart failure, would result in lower power efficiency of the cardiorespiratory system.  This argument is validated by the increases in cardiac output observed in multiple animal models of heart failure paced with artificially restored RSA~\cite{OCallaghan2019,Shanks2022}.  In healthy subjects, the functional significance of RSA may thus be to save the heart energy.  Now commenting on the quantitative discrepancy between the prediction of a $10\%$ increase in cardiac efficiency and the experimentally observed $17-10\%$ increase in cardiac output, we note that our argument focused on the pulmonary circuit omitting wider effects of RSA on the systemic circuit.  Other coupling mechanisms will act in parallel with viscoelastic coupling.  For example RSA is known to also improve pulmonary gas exchange and circulatory efficiency~\cite{Hayano1996}.  Ben Tal et al.~\cite{BenTal2012} have argued that the purpose of RSA is to minimize the work done by the heart while maintaining physiological levels of arterial carbon dioxide.  Validating the $10\%$ efficiency gain \textit{in-vivo} would require measuring the effect of dissipation-synchronization independently from other coupling mechanisms which might prove challenging.  At this stage, the broad agreement of our predictions with experimentally observed efficiency gains (within a factor 2) makes our argument plausible even on a quantitative level.

\textit{In-vivo} experiments~\cite{OCallaghan2019,Shanks2022} used natural breathing to drive RSA modulation.  The breathing cycle was thus aperiodic and varied over the sleep / wake cycle.  We have simulated aperiodic breathing (Supplementary information 4) and find that cardiorespiratory synchronization is robust as long as the standard deviation of breathing intervals is less than the width of Arnold tongues.  One therefore expects that the efficiency gains induced by RSA will be sustained under mildly aperiodic breathing.

One difference between natural RSA and the RSA produced by the present model is that the RSA dose was designed to be set with the $rsa$ parameter to meet a therapeutic need.  In contrast, natural RSA produces a cardiac modulation that is largest in the resting state and decreases as the heart rate increases~\cite{Bartsch2012}.  The constant RSA dose produced by our model does not however constitute a shortcoming because patients in need of RSA therapy are invariably in a state of fatigue and breathlessness, that corresponds to the resting state and where RSA pacing is most beneficial.  Bartsch et al~\cite{Bartsch2012} have further reported that the strength of cardiorespiratory synchronization is independent of breathing frequency.  Our model does fully reproduce this behaviour as the coupling strength is set by the $rsa$ parameter hence is independent of breathing frequency.  Our model can further explain the observation of different $m:n$ synchronization modes across different sleep states~\cite{Bartsch2007}.  Transitions between the $3:1$ and $4:1$ modes of Fig.\ref{fig:fig3} would easily occur following a $26\%$ change in heart rate between normal and REM sleep~\cite{Bartsch2007}.

It is useful to note that passive synchronization by viscoelastic damping is almost certainly present in the cardiorespiratory system.  This is a purely mechanical effect. It would subsist if central nervous control were lost.  Passive synchronization is expected to occur because the pulmonary vasculature dissipates mechanical energy at a rate greater than the frequency difference between two stable cardiorespiratory modes, either $3:1$ and $4:1$ or $3:1$ and $2:1$ ~\cite{Winfree1967,Lu2023,Hale1996,Masson2022,Zhu2024,Cabot2020}.  Taking $\eta=25-30$kg.m$^{-1}$.s$^{-1}$ as the viscosity of lung tissue~\cite{Nordsletten2021,Suki2021} and $E=70-260$kPa as its Young's modulus, the viscoelastic damping rate is $E/\eta=2.3-10.4$s$^{-1}$.  In comparison, the frequency detuning of the $3:1$ mode relative to either $2:1$ or $4:1$ is $|\omega_{4:1}-\omega_{3:1}|=|\omega_{2:1}-\omega_{3:1}|=0.33$s$^{-1}$. The damping rate being faster than the frequency detuning, the amplitude of anti-phasic oscillations of the heart-lung pair will be rapidly damped leaving only phasic oscillations in the steady state~\cite{Lu2023}.  The heart-lung system therefore invariably evolves towards the synchronized state.  This mechanism is important as it can explain why cardiorespiratory synchronization is still observed in heart transplant patients who have lost central nervous system control~\cite{Mortara1994,Toledo2001}.

An alternative interpretation involving cardio-ventilatory coupling has been hypothesized~\cite{Larsen1999,Gallety1997} to explain cardiorespiratory synchronization in heart transplant patients.  This posits a mechanism, converse to RSA, whereby cardiac oscillations control the timing of inspiration.  By periodically modulating oxygen demand relative to a threshold, heart beats may trigger a breadth intake once a oxygen saturation threshold is crossed~\cite{McGuinness2004}.  Such integrate-and-fire mechanism can reset synchronization~\cite{Glass2001}, albeit sparsely, as it is only activated once every tens of respiratory cycles.  This is because the mean arterial $O_2$ (and $CO_2$) partial pressure varies very slowly, on the scale of minutes, instead of seconds in the case of the respiratory cycle.  Hence the saturation threshold would be attained after tens of breathing cycles producing sparse breathing events to correct arterial gas imbalances~\cite{Gallety1997}.  In contrast synchronization-dissipation acts on a beat-to-beat basis.  For this reason it will produce a more effective coupling than cardio-ventilatory coupling.  In addition, it arguably provides a simpler explanation to phasic synchronization between cardiac and respiratory rhythms~\cite{Schafer1999,Rzeczinski2002}.

The synchronization-dissipation principles described here could be inserted in large-scale integrative simulations~\cite{Sarmiento2021} to understand the full physiological import of nonlinear and dissipative cardiorespiratory coupling.

\section{Conclusions}

We have modelled the interactions between the cardiac and the respiratory systems through the pulmonary vascular impedance when the timing of cardiac pulses relative to inspiration is controlled by a model of the central nervous system.  The entrainment of the cardiac oscillator by breathing modulates heart rate (RSA) and produces phasic synchronization when the ratio of the mean cardiac to respiratory frequency is a rational number ($n:m$).  Within domains of synchronization, viscoelastic losses in the pulmonary vasculature are reduced by as much as $55\%$.  At the $3:1$ cardiac-to-breathing frequency ratio relevant to humans at rest, viscoelastic dissipation is $\approx 10\%$ lower under RSA pacing.  This increases the pacing efficiency of the right ventricle by a corresponding amount.  Because these results are consistent with $\textit{in-vivo}$ data, it is surmised that one function of RSA may be to improve right ventricle function and overall cardiac pumping efficiency.

\section{Author Contributions}

JB computed the viscoelastic power dissipation maps.  AL and VJ computed Arnold tongue plots.  AN planned the work, built the model, checked simulations, and wrote the manuscript.  All authors approved the final version of the manuscript.
\\

\section{Acknowledgements}

A.N. wishes to thank Profs Julian Paton, Mariann Gy{\"o}ngy{\"o}si and Marc Vos for many illuminating discussions on the genesis of RSA and cardiac electrophysiology and Mr Ben Walker for code checking.
\\

\section{Financial disclosure}

This research was supported by the European Union under the Horizon2020, Future Emerging Technologies project 732170 and the European Innovation Council under Transition project 101214165.

\bibliography{Biblio_Pace_dissipation}

\end{document}